# Theory of resonance energy transfer involving nanocrystals: the role of high multipoles

### Roi Baer
*Department of Physical Chemistry and the Fritz Haber Center for Molecular Dynamics, the Hebrew University of Jerusalem, Jerusalem 91904 Israel.*

### Eran Rabani
*School of Chemistry, the Raymond and Beverly Sackler Faculty of Exact Sciences, Tel Aviv University, Tel Aviv 69978 Israel.*

A theory for the fluorescence resonance energy transfer (FRET) between a pair of semiconducting nanocrystal quantum dots is developed. Two types of donor-acceptor couplings for the FRET rate are described: dipole-dipole (d-d) and the dipole-quadrupole (d-q) coupling. The theory builds on a simple effective mass model which is used to relate the FRET rate to measureable quantities such as the nanocrystal size, fundamental gap, effective mass, exciton radius and dielectric constant. We discuss the relative contribution to the FRET rate of the different multipole terms, the role of strong to weak confinement limits, and the effects of nanocrystal sizes.

## I. INTRODUCTION

The development of novel sensing, imaging and biological labeling is an expanding research field in recent years.[1-6] In particular, fluorescence probes are widely used in single molecule imaging[1,2] and spectroscopy,[3] and in the detection techniques of proteins, peptides and enzymes.[4-6] Early studies were based mainly on organic dye molecules as fluorophores. However, since organic dyes have very broad emission lines and fast photobleaching, their applications are quite limited. More recently, semiconductor nanocrystal quantum-dots (QDs) have been suggested as potential fluorophores.[7-14] The nanocrystal QDs exhibit very narrow emission bands that can be tuned by simply changing the size or composition of the nanocrystals, thus providing simple means to control the probe properties. Due to their brightness (and also low photobleaching) very low light intensity can be used, practical for many biological applications (in particular for living cells). Furthermore, their wide absorption band allows simultaneous excitation of several different probes, providing new directions in fluorescence probing.

One of the more common fluorescence techniques for probing biological systems is based on FRET between a donor and an acceptor.[15] For example, studies based on FRET have been used to probe structural changes in protein conformations.[3] In principle, FRET is a sensitive tool for studying the separation between the donor and the acceptor, providing structural information in real-time. In this respect, semiconductor nanocrystals offer an additional advantage over organic dyes – their size can be tuned and thus different "rulers" can be used ranging from several angstroms to several nanometers.

There is an important synergism between experiments and theory in the study of FRET. At the heart is the mapping between the experimental measured FRET signal and the distance between the donor and acceptor. Typically, this is established through the FRET efficiency defined as $\epsilon^{-1} = 1 + \tau/k_{DA}$, where $k_{DA}$ is the FRET rate and $\tau$ is the fluorescence lifetime. The common approach taken for molecular donor/acceptor systems is based on Förster resonance energy transfer theory, where nonradiative energy transfer from an excited donor molecule to an acceptor molecule takes place.[15] Based on second order perturbation theory (Fermi's Golden Rule) combined with the lowest order multipole expansion of the transition moments of the donor and acceptor, Förster showed that the FRET rate depends on the center-to-center separation between the donor and acceptor, $R$, and scales as $k_{DA} \propto R^{-6}$.

The application of Förster theory to the case where the probes involve semiconductor nanocrystal QDs is highly questionable. As pointed out in Ref. [8] "The Förster theory treats the donor and acceptor as points in the interaction space, whereas the nanocrystals have finite size and are relatively large compared to the dye molecules. Nonetheless, this treatment is the best available for the present scenario." The multipole expansion of transition moments is expected to break down on length-scales comparable to nanocrystal size, exactly the lengths probed by FRET experiments. The Förster theory has been extended in several different directions including the case of higher multipoles and short-range effects.(see Ref. [16] and references therein). However, the application of these modified theories to semiconducting nanocrystal QDs is still quite limited and involves hard-core simulations where the simplicity of the Förster theory is lost.

In the present study we extend the Förster theory to treat the case where the donor, or acceptor, or both, are semiconductor nanocrystal QDs. Unlike previous work,[17,18] the present approach explicitly treats the electronic structure of the nanocrystals, adopting a simple model based on the effective mass approximation. This approach does not take into account electron spin coupling, crystal fields, electron-hole exchange interactions and inter-band couplings. More accurate treatments based on a Luttinger multi-band $\mathbf{k} \cdot \mathbf{p}$ model[19,20] or on a semiempirical atomistic treatment[21-26] will be the subject of future study. Within the simple effective mass model, we retain the spherical symmetry of the QD and treat the effects of higher multipoles on the FRET rate.[27-29] Both weak and strong confinement limits are discussed. In the former case, approximate expressions for the distance dependent FRET rate including dipole and quadrupole transition moments are derived. Examples are given for realistic model parameters of CdSe nanocrystals.

## II. RESONANT ENERGY TRANSFER THEORY

Consider a donor and acceptor in a medium of dielectric constant $\varepsilon$. The rate of energy transfer from donor to acceptor in FRET theory is given by the Fermi golden rule expression:

$$W_{D \to A} = \frac{2\pi}{\hbar} \sum_{if} \rho(i) |V_{if}|^2 \delta(E_i - E_f) \quad (2.1)$$

Here, $|i\rangle = |0\delta\rangle$ is the initial state, where the donor is in an excited state $\delta$ and the acceptor is in the ground state 0, and $|f\rangle = |0\alpha\rangle$ is the final state where the excitation was transferred to the acceptor at state $\alpha$. $\delta(E_i - E_f)$ ensures conservation of energy between initial and final states, $\rho(i)$ is the probability of having the initial state $|i,0\rangle$. In the above, $V_{if}$ is the matrix element of the electromagnetic coupling between the states which can be expressed as a multipole expansion around a central points of the donor and acceptor separated by a vector $\mathbf{R}$:

$$V_{if} = \left(d_n^\delta \vec{\partial}_n + \hat{\Theta}_{nm}^\delta \vec{\partial}_n \vec{\partial}_m + ...\right)\frac{1}{\varepsilon R} \\ \left(d_{n'}^\alpha \vec{\partial}_{n'} + \hat{\Theta}_{n'm'}^\alpha \vec{\partial}_{n'} \vec{\partial}_{m'} + ...\right) \quad (2.2)$$

Where $d_n^\delta = \langle 0|e\hat{r}_n|\delta\rangle$ is the $n$ Cartesian component of the donor transition dipole moment ($\hat{\mathbf{r}}$ is the position vector of an electron) and $\Theta_{nm}^\delta$ is the $n,m$ component of the transition quadrupole moment ($n,m = x,y,z$). Similar definitions apply for the acceptor (with $\delta \to \alpha$). Einstein summation convention is used.

The two lowest order terms of $V_{if}$ considered in this work are the dipole-dipole (d-d) and dipole-quadrupole (d-q) terms. The dipole-dipole coupling matrix element is given by:

$$V_{if}^{d-d} = \frac{R^2 (\mathbf{d}^\delta \cdot \mathbf{d}^\alpha) - 3(\mathbf{d}^\delta \cdot \mathbf{R})(\mathbf{d}^\alpha \cdot \mathbf{R})}{\varepsilon R^5} \\ = \frac{d^\delta d^\alpha}{\varepsilon R^3}\left(\cos\theta_{\alpha\delta} - 3\cos\theta_\delta \cos\theta_\alpha\right) \quad (2.3)$$

Where $\theta_{\alpha\delta}$ is the angle between the transition dipole moments and $\theta_\delta$ ($\theta_\alpha$) is the angle between $\mathbf{d}^\delta$ ($\mathbf{d}^\alpha$) and $\mathbf{R}$ (see sketch in Figure 1). Similarly, for the dipole-quadrupole interaction:

$$V_{if}^{d-q} = \frac{3}{\varepsilon R^7}\left\{R^2\left((\mathbf{d}^\delta \cdot \mathbf{R}) tr\vec{\Theta}^\alpha + 2\mathbf{d}^\delta \cdot \vec{\vec{\Theta}}^\alpha \cdot \mathbf{R}\right) \\ - 5(\mathbf{d}^\delta \cdot \mathbf{R})(\mathbf{R} \cdot \vec{\vec{\Theta}}^\alpha \cdot \mathbf{R})\right\} \\ = \frac{3d^\delta \Theta^\alpha}{\varepsilon R^4}\left(\sin\theta_\delta \sin 2\theta_\alpha \cos(\phi_\delta - \phi_\alpha) \\ + \cos\theta_\delta (1 - 3\cos^2\theta_\alpha)\right) \quad (2.4)$$

The second equality in Eq. (2.4) assumes that the transition quadrupole moment is a spherically symmetric tensor given by:

$$\vec{\vec{\Theta}}^\alpha = \Theta^\alpha \begin{pmatrix} \sin^2\theta_\alpha \cos^2\phi_\alpha & \frac{1}{2}\sin^2\theta_\alpha \sin 2\phi_\alpha & \frac{1}{2}\sin 2\theta_\alpha \cos\phi_\alpha \\ \frac{1}{2}\sin^2\theta_\alpha \sin 2\phi_\alpha & \sin^2\theta_\alpha \sin^2\phi_\alpha & \frac{1}{2}\sin 2\theta_\alpha \sin\phi_\alpha \\ \frac{1}{2}\sin 2\theta_\alpha \cos\phi_\alpha & \frac{1}{2}\sin 2\theta_\alpha \sin\phi_\alpha & \cos^2\theta_\alpha \end{pmatrix} \quad (2.5)$$

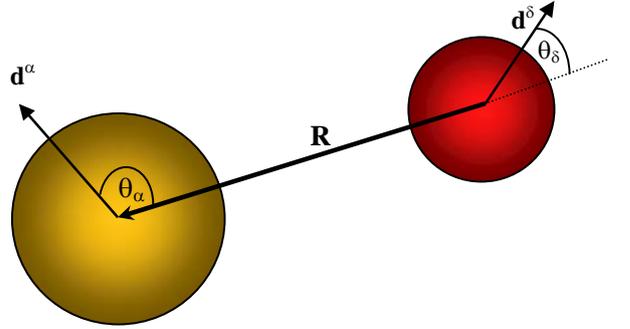

Figure 1: A sketch of the transitions dipoles between donor (small dot) and acceptor (large dot). $\mathbf{R}$ is directed taken along the $z$ axis.

Similarly to the dipole case, the transition quadrupole moment is written as a spherical quantity and a canonical angular dependence, depending on $\theta_\alpha$ and $\phi_\alpha$. Similar expressions exist also for quadrupole-dipole coupling and other terms not described here. The contribution of these terms to the FRET rate can be neglected as will be discussed below.

The treatment above neglects polarization effects inside the QD arising from the fact that it has a different dielectric constant $\varepsilon_{QD}$ than the surrounding medium. An approximate simple way to account for $\varepsilon_{QD}$ is to multiply the $\ell$-order term in the multipole expansion by a local field factor (see Appendix A for details):

$$f_{\ell,D/A} = \frac{2\ell + 1}{(\varepsilon_{QD}/\varepsilon + 1)\ell + 1}, \quad (2.6)$$

Thus the local field factor for the QD dipole moment is $f_1 = 3/(2 + \varepsilon_{QD}/\varepsilon)$ and for the quadrupole moment is $f_2 = 2.5/(1.5 + \varepsilon_{QD}/\varepsilon)$. For aqueous solution ($\varepsilon \approx 80$) both factors are of the order of $\frac{3}{2}$. For organic solvents ($\varepsilon \approx 2$) they are on the order of $\frac{1}{2}$.

In most applications of FRET is it custom to perform an average of the FRET rate over the angles ($\theta_\delta, \theta_\alpha, \phi_\delta$ and $\phi_\alpha$). This is a consequence of the fact that the sample is heterogeneous or in some cases due to a long time self-averaging mechanism. The details are given in Appendix B. For the dipole-dipole and dipole-quadrupole rates we obtain:

$$W_{D \to A}^{d-d} = \frac{2}{3} \frac{2\pi}{\hbar \varepsilon^2} \frac{f_{1,D}^2 f_{1,A}^2}{R^6} \sum_{\alpha\delta} \rho(\delta) d_\delta^2 d_\alpha^2 \delta(\varepsilon_\delta - \varepsilon_\alpha)$$
$$W_{D \to A}^{d-q} = 4 \frac{2\pi}{\hbar \varepsilon^2} \frac{f_{1,D}^2 f_{2,A}^2}{R^8} \sum_{\alpha\delta} \rho(\delta) d_\delta^2 \Theta_\alpha^2 \delta(\varepsilon_\delta - \varepsilon_\alpha) \quad (2.7)$$

The sum over $\alpha$ ($\delta$) runs over all states of the acceptor (donor). Decomposing the $\delta$-function as

$$\delta(\varepsilon_\delta - \varepsilon_\alpha) = \int \delta(\varepsilon - \varepsilon_\alpha) \delta(\varepsilon_\delta - \varepsilon) d\varepsilon, \quad (2.8)$$

We arrive at the final expressions for the FRET rates, given in terms of spectral overlap integrals:

$$W_{D \to A}^{d-d} = \frac{4\pi}{3\hbar \varepsilon^2} \frac{f_{1,D}^2 f_{1,A}^2}{R^6} \int D_{dip}(\varepsilon) A_{dip}(\varepsilon) d\varepsilon$$
$$W_{D \to A}^{d-q} = \frac{8\pi}{\hbar \varepsilon^2} \frac{f_{1,D}^2 f_{2,A}^2}{R^8} \int D_{dip}(\varepsilon) A_{quad}(\varepsilon) d\varepsilon \quad (2.9)$$

Where the emission $(D)$ and absorption $(A)$ spectral functions are:

$$D_{dip}(\epsilon) = \sum_\delta \rho(\delta) d_\delta^2 \delta(\epsilon - \epsilon_\delta)$$
$$A_{dip}(\epsilon) = \sum_\alpha d_\alpha^2 \delta(\epsilon - \epsilon_\alpha) \quad (2.10)$$
$$A_{quad}(\epsilon) = \sum_\alpha \Theta_\alpha^2 \delta(\epsilon - \epsilon_\alpha)$$

As expected, the dependence of the FRET rates on the distance between donor and acceptor is different for the dipole-dipole and dipole-quadrupole interaction. When $R$ is large compared to the particle sizes the dipole-dipole term dominates the overall FRET rate. The situation becomes more complicated when $R$ is comparable to the system sizes, where a close examination of the spectral functions is required, as discussed below in Section IV.

The FRET rate is essentially an overlap between the spectral functions of the donor and acceptor. To relate it to measurable quantities, we note that the spectral functions are related to the absorption cross-sections and the normalized emission spectra:[27]

$$\bar{D}_{em}(\omega) = \frac{1}{2\pi c\tau} \left(\frac{\omega}{c}\right)^3 \frac{4}{3} D_{dip}(\hbar\omega)$$
$$A_{dip-abs}(\omega) = \frac{(2\pi)^2 N_A}{3000 \ln 10} \frac{\omega}{c} A_{dip}(\hbar\omega) \quad , \quad (2.11)$$
$$A_{quad-abs}(\omega) = \frac{(2\pi)^2 N_A}{6000 \ln 10} \left(\frac{\omega}{c}\right)^3 A_{quad}(\hbar\omega)$$

where the normalized spectrum obeys:

$$2\pi c \int \bar{D}_{em}(2\pi c \bar{\nu}) d\bar{\nu} = 1 \quad (2.12)$$

and $\bar{\nu}$ is the inverse wavelength. For an isolated transition at $\bar{\nu}$ a relation between the radiative relaxation time $\tau$ and the integrated dipole spectral function can be obtained:

$$\frac{1}{\tau} = \frac{\phi_D}{\tau_D} = \int \frac{4\omega^3}{c^3} D_{dip}(\hbar\omega) d\omega \quad (2.13)$$

where, $\phi_D$ is the quantum yield, and $\tau_D$ is the total life-time of the donor excited state. The radiative lifetime is also connected to the transition dipole moment:

$$d_B^2 = \frac{\hbar}{4} \frac{\phi_D}{\tau_D (2\pi\bar{\nu})^3} \quad (2.14)$$

In terms of these experimental measureable quantities, the FRET rate is given by:

$$W_{D \to A}^{d-d} = \frac{2}{3} \frac{\phi_D C_7}{\tau_D \varepsilon^2} \frac{f_{1,D}^2 f_{1,A}^2}{R^6} \int \bar{D}_{em}(2\pi c\bar{\nu}) A_{dip-abs}(2\pi c\bar{\nu}) \frac{d\bar{\nu}}{\bar{\nu}^4}$$
$$W_{D \to A}^{d-q} = 4 \frac{\phi_D C_9}{\tau_D \varepsilon^2} \frac{f_{1,D}^2 f_{2,A}^2}{R^8} \int \bar{D}_{em}(2\pi c\bar{\nu}) A_{quad-abs}(2\pi c\bar{\nu}) \frac{d\bar{\nu}}{\bar{\nu}^6}$$
(2.15)

where:

$$C_k = \frac{9000 \ln 10}{2^k \pi^{k-2} N_A}, \qquad k = 7, 9 \quad (2.16)$$

The first equation in (2.15) is the Förster formula.[16] One of the fundamental consequences of Eqs. (2.15) is that one can take into account inhomogenous effects simply by using the inhomogeneous broadened spectra of the donor and acceptor.

## III. EFFECTIVE MASS MODEL

To calculate the FRET rate given by Eq. (2.15) one requires as input the transition multipole moments and the energy spectrum of the donor and acceptor. Here, we adopt the effective mass model to describe these properties for the nanocrystal QDs. This model does not describe the excitonic fine structure and in particular the bright and dark states,[19] however, it captures some of the spectral features of nanocrystals[30-35] and facilitates the analysis of size dependence and other parameters. A more realistic treatment of the electronic structure based on a Luttinger multi-band $\mathbf{k}\cdot\mathbf{p}$ model[19,20] or on a semiempirical atomistic treatment[21-26] will be used for calculating the FRET rate of nanocrystals in a future publication.

### A. The electron-hole wave functions and energies

We consider a 2 band (valence and conductance) system. The eigenfunctions of the holes and electrons are written as a product of an envelope function $\phi_{e,h}(\mathbf{r})$ and a lattice periodic function $u_{V,C}(\mathbf{r})$:

$$\begin{aligned}\varphi_h(\mathbf{r}) &= \phi_h(\mathbf{r})u_V(\mathbf{r})\\ \varphi_e(\mathbf{r}) &= \phi_e(\mathbf{r})u_C(\mathbf{r})\end{aligned} \qquad (3.1)$$

The envelope functions are the zero order eigenfunctions of the electron-hole pair Hamiltonian:

$$\hat{H} = -\frac{\hbar^2}{2m_e}\nabla_e^2 - \frac{\hbar^2}{2m_h}\nabla_h^2 + V_{conf}(\mathbf{r}_e,\mathbf{r}_h) - \frac{e^2}{\varepsilon_{QD}|\mathbf{r}_e - \mathbf{r}_h|}, (3.2)$$

where $m_e$ ($m_h$) is the electron (hole) effective mass, $\varepsilon_{QD}$ is the dielectric constant of the nanocrystal, and the last term on the right hand side is the perturbation term. For a spherical quantum dot, the confinement potential is taken as 0 inside the dot and $\infty$ outside. The orthonormal envelope functions are then zero outside of the dot and inside it are given by:

$$\phi_{nlm}(\mathbf{r}) = N_{n,l} j_l\left(\kappa_{n,l}\frac{r}{R_{QD}}\right)Y_{lm}(\theta,\phi); \quad r < R_{QD}, \qquad (3.3)$$

Where $Y_{lm}(\theta,\phi)$ is a spherical harmonics function, $j_l(x)$ is a spherical Bessel function with $\kappa_{n,l}$ its $n^{th}$ zero, and $R_{QD}$ is the nanocrystal radius and $N_{n,l} = \sqrt{\frac{2}{R^3}}(j_{l+1}(\kappa))^{-1}$ is the normalization constant. Note, that in this model the envelope functions for the hole and the electron are identical (since $\kappa_{nl}$ is independent of the effective mass). This form neglects the electron-hole pair interaction which is included to first order in the energies only, given by:[30-32]

$$E_{nl,n'l'} = E_g + \frac{\hbar^2 \kappa_{nl}^2}{2m_e R_{QD}^2} + \frac{\hbar^2 \kappa_{n'l'}^2}{2m_h R_{QD}^2} - \frac{1.8e^2}{\varepsilon_{QD} R_{QD}}. \qquad (3.4)$$

Here, $E_g$ is the bulk band gap and the last term represents the electron-hole interaction to first order assuming spherical symmetric wavefunctions for both the electron and the hole.

### B. The Transition Moments

To calculate the transition multipoles, note two properties of the integrals concerning the envelope and lattice periodic functions:

$$\begin{aligned}\langle\phi_e|\phi_h\rangle_{space} &\approx \Omega\sum_{\mathbf{L}}\phi_h(\mathbf{r}+\mathbf{L})\phi_e(\mathbf{r}+\mathbf{L})\\ \langle u_C|u_V\rangle_{cell} &= 0\end{aligned}. \qquad (3.5)$$

The first approximation is a result of the slow variations of the envelope functions on a scale of $\Omega^{1/3}$, where $\Omega$ is the volume of a unit cell. The second equality is a property of the Bloch functions in the bulk.

Using Eq. (3.5), the transition dipole moment $\mathbf{d}$ is given by the product of the envelope overlap and the unit cell bulk transition dipole:

$$\begin{aligned}\mathbf{d} &= e\int_{\text{all space}}\varphi_h(\mathbf{r})\mathbf{r}\varphi_e(\mathbf{r})d^3r\\ &\approx \langle\phi_h|\phi_e\rangle\frac{e}{\Omega}\int_\Omega u_V(\mathbf{r})\mathbf{r}u_C(\mathbf{r})d^3r\end{aligned}, \qquad (3.6)$$

Within the simple effective mass model and infinite confinement potential the overlap between the electron and hole wave functions is $\langle\phi_h|\phi_e\rangle = \delta_{eh}$ since $\phi_e$ and $\phi_h$ are independent of the effective mass of the electron and hole. This implies that dipole allowed transitions occur only when $n = n'$, $l = l'$ and $m = m'$, i.e. from the ground state to $1S_e - 1S_h$, $1P_e - 1P_h$, etc. In this case

$$\mathbf{d} \approx \frac{e}{\Omega}\langle u_V|\mathbf{r}|u_C\rangle_\Omega \delta_{eh} = \mathbf{d}_B \delta_{eh}, \qquad (3.7)$$

where $\mathbf{d}_B$ is the bulk transition dipole moment. The transition dipole moment is essentially equal to the transition dipole moment of the bulk unit cell and is independent of the size of the nanocrystal QD.[36]

We note in passing that for a finite confinement potential,[37] or when the electron and hole interactions are treated exactly,[33] or for a more elaborate model,[19,38,39] one needs to calculate the envelope overlap which may depend on the nanocrystal effective masses and size.

The quadrupole transition moment involves the bilinear product $\overset{\leftrightarrow}{\Theta} = e\,\mathbf{r}\mathbf{r}^T$, given by the integral:

$$\begin{aligned}(\overset{\leftrightarrow}{\Theta})_{he} = e\sum_{\mathbf{L}}\int_\Omega \phi_h(\mathbf{r}+\mathbf{L})u_V(\mathbf{r})\times\\ \times(\mathbf{r}+\mathbf{L})(\mathbf{r}+\mathbf{L})^T\phi_e(\mathbf{r}+\mathbf{L})u_C(\mathbf{r})d^3r\end{aligned}. \qquad (3.8)$$

It decomposes into four contributions:

$$\left(\ddot{\Theta}\right)_{he} = e\sum_{\mathbf{L}}\int_{\Omega}\phi_h\left(\mathbf{r}+\mathbf{L}\right)u_V\left(\mathbf{r}\right)\mathbf{r}\mathbf{r}^T\phi_e\left(\mathbf{r}+\mathbf{L}\right)u_C\left(\mathbf{r}\right)d^3r$$
$$+e\sum_{\mathbf{L}}\int_{\Omega}\phi_h\left(\mathbf{r}+\mathbf{L}\right)u_V\left(\mathbf{r}\right)\mathbf{L}\mathbf{r}^T\phi_e\left(\mathbf{r}+\mathbf{L}\right)u_C\left(\mathbf{r}\right)d^3r$$
$$+e\sum_{\mathbf{L}}\int_{\Omega}\phi_h\left(\mathbf{r}+\mathbf{L}\right)u_V\left(\mathbf{r}\right)\mathbf{r}\mathbf{L}^T\phi_e\left(\mathbf{r}+\mathbf{L}\right)u_C\left(\mathbf{r}\right)d^3r \quad (3.9)$$
$$+e\sum_{\mathbf{L}}\mathbf{L}\mathbf{L}^T\int_{\Omega}\phi_h\left(\mathbf{r}+\mathbf{L}\right)u_V\left(\mathbf{r}\right)\phi_e\left(\mathbf{r}+\mathbf{L}\right)u_C\left(\mathbf{r}\right)d^3r$$

The first term involves the unit cell transition quadrupole which is does not depend on the QD size and therefore, is neglected. The last term involves the overlap of the conduction and valence lattice periodic functions and is thus zero. Finally, the second and third terms evaluate to:

$$e\left(\ddot{\Theta}\right)_{he} = \mathbf{d}_B \mathbf{d}_{env}^T + \mathbf{d}_{env}\mathbf{d}_B^T \quad (3.10)$$

Where $\mathbf{d}_B$ is defined in Eq. (3.7) and $\mathbf{d}_{env}$ is proportional to the QD radius and is given by:

$$\mathbf{d}_{env} = e\left\langle\phi_e\left|\mathbf{r}\right|\phi_h\right\rangle \quad (3.11)$$

As far as we know, the integral in Eq. (3.11) has no exact analytical solution, despite its apparent simplicity. For the $z$ component of the envelope dipole we found that only states with $m = m'$ and $l = l' \pm 1$ are allowed. Furthermore, numerically it is found that states with $n = n'$ and $n = n' \pm 1$ have significantly higher envelope transition moment than other combinations. Thus, a reasonable approximation for the $z$ component envelope transition moment is given by (see Appendix C for further details):

$$\left\langle\phi_{h,nlm}\left|z\right|\phi_{e,n'l'm'}\right\rangle = \frac{R_{QD}}{2} \times$$
$$\left\{F_{l+1}^m\delta_{l+1,l'}\left(\delta_{n,n'}+\frac{2}{3}\delta_{n',n-1}\right)\right. \quad (3.12)$$
$$\left.+F_l^m\delta_{l-1,l'}\left(\delta_{n,n'}+\frac{2}{3}\delta_{n',n+1}\right)\right\}$$

Similar results can be obtained for the $x$ and $y$ components with $m = m' \pm 1$.

### C. The absorption spectral functions

Incorporating into Eq. (2.10) the energy levels given by Eq. (3.4) and the selection rules for the dipole transition (Eq. (3.7)), the dipole spectral function is then given by:[36]

$$A_{dip}\left(\varepsilon\right) = d_B^2 \sum_{nl}\delta\left(E_g + \frac{\hbar^2\kappa_{nl}^2}{2\mu R_{QD}^2} - \frac{1.8e^2}{\varepsilon_{QD}R_{QD}} - \varepsilon\right)(2l+1), \quad (3.13)$$

where $\mu$ is the reduced electron-hole effective mass ($\mu^{-1} = m_e^{-1} + m_h^{-1}$).

In the weak confinement limit ($R_{QD} \gg a_B$ where $a_B = \hbar^2\epsilon_{QD}/\mu e^2$ is the Bohr radius for the electron-hole pair),[36] we can replace the sum by an integral over the continuous spectrum of a particle in a large sphere to obtain:

$$A_{dip}\left(\varepsilon\right) \approx \frac{d_B^2}{3\pi}\frac{2\mu}{\hbar^2}R_{QD}^3\sqrt{\frac{2\mu}{\hbar^2}\left(\varepsilon - E_g + \frac{1.8e^2}{\varepsilon_{QD}R_{QD}}\right)} \quad (3.14)$$
(weak confinement)

This result shows that the dipole spectral function scales approximately as $R_{QD}^\alpha$ where $\alpha \approx 2.5 - 3$, consistent with the experimental observation for absorption spectrum of nanocrystals QDs.[40,41]

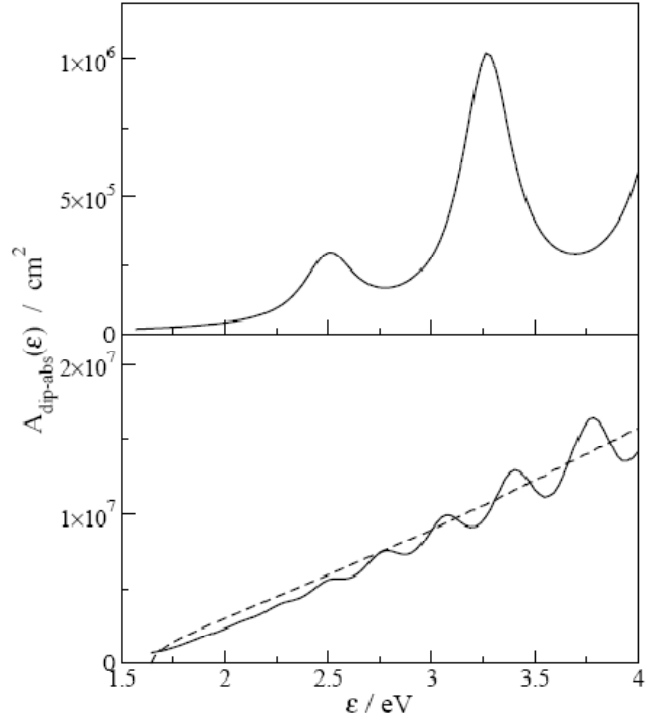

Figure 2: The dipole absorption cross-section of a 2 nm (upper panel) and 5 nm (lower panel) CdSe nanocrystal. Effective masses (in atomic units): $m_e = 0.13$ and $m_h = 0.45$, $E_g = 1.7\ eV$, $\tau = 1\,ns$ ($d_B \approx 20\,Debeye$) and $\varepsilon_{QD} = 10$. The broadening parameter is $\eta = 0.05$ eV. The dashed line in the lower panel is the weak confinement approximation given by Eq. (3.14) combined with (2.11).

In many situations, the case of the strong confinement is more relevant as the discrete nature of the exciton states becomes important. In this limit, one is required to perform the sum given by Eq. (2.10) for to calculate the dipole spectral function. Often, only very few exciton states lie within the relevant energy range and thus the sum can be represented by a small number of terms. To better represent the spectrum in this strong confinement limit, we include a broadening of the $\delta$ - functions by a Lorentzian profile:

$$\delta\left(\varepsilon\right) \to \frac{1}{\pi}\text{Im}\frac{1}{\varepsilon - i\eta} \quad (3.15)$$

where $\eta$ is the energy broadening parameter. The actual value of the broadening parameter depends on the type of measurement one makes. A reasonable value for a inhomogeneous broadened spectrum is $\eta = 0.15$ eV .[40-42]

In Figure 2 we plot the dipole absorption spectrum as given by the combination of Eqs. (2.11) and (3.13) of a CdSe nanocrystal in the strong (upper panel, $R_{QD} = 2nm$) and weak (lower panel, $R_{QD} = 5nm$) confinement limits. In the strong confinement limit we find that the lowest transition observed is to the $1S_e - 1S_h$ exciton state at 2.5 eV, followed by the transition to the $1P_e - 1P_h$ state at 3.3 eV. In the lower panel of Figure 2 we compare the exact numerical result given by Eq. (3.13) to the weak confinement limit approximation of Eq. (3.14). For the relevant energy regime the approximation captures the essential behavior of the spectrum.

The quadrupole spectral function is obtained by incorporating into Eq.(2.10) the energy levels given by Eq. (3.4) and the selection rules for the quadrupole transition given by Eq.(3.10). For simplicity, we consider only the $zz$ component of the quadrupole moment. Using Eq. (3.10) with Eq.(3.12), we obtain after some algebra:

$$A_{quad}(\varepsilon) = \frac{1}{3}\left(R_{QD}d_B\right)^2 \times$$
$$\sum_{nl}(l+1)\left\{\delta\left(\Delta E_{nl}^{n(l+1)} - \varepsilon\right) + \frac{4}{9}\delta\left(\Delta E_{(n+1)l}^{n(l+1)} - \varepsilon\right)\right.$$
$$\left. + \delta\left(\Delta E_{n(l+1)}^{nl} - \varepsilon\right) + \frac{4}{9}\delta\left(\Delta E_{n(l+1)}^{(n+1)l} - \varepsilon\right)\right\} \quad (3.16)$$

where:

$$\Delta E_{nl}^{n'l'} = E_g - \frac{1.8e^2}{\varepsilon_{QD}R_{QD}} + \frac{\hbar^2}{2\mu R_{QD}^2}\left(\frac{\gamma\kappa_{nl}^2 + \kappa_{n'l'}^2}{1+\gamma}\right), \quad (3.17)$$

and $\gamma = m_e/m_h$. In deriving Eq. (3.16) we have also used the relations $\sum_{m=-l}^{l} F_{lm}^2 = \frac{l}{3}$ and $\sum_{m=-l}^{l} F_{(l+1)m}^2 = \frac{l+1}{3}$.

In the weak confinement limit we assume that the separation between quadrupole allowed transitions is small and thus the broadened $\delta$-functions in Eq. (3.16) are nearly overlapping. This leads to a simplified expression for the quadrupole spectrum given by:

$$A_{quad}(\varepsilon) \approx \left(R_{QD}d_B\right)^2 \sum_{nl}(l+1)\delta\left(\Delta E_{nl}^{nl} - \varepsilon\right)$$
$$\approx \frac{1}{2}R_{QD}^2 A_{dip}(\varepsilon) \quad (3.18)$$
$$\approx \frac{d_B^2}{3\pi}\frac{\mu}{\hbar^2}R_{QD}^5\sqrt{\frac{2\mu}{\hbar^2}\left(\varepsilon - E_g + \frac{1.8e^2}{\varepsilon_{QD}R_{QD}}\right)}$$
(weak confinement)

In Figure 3 we plot the quadrupole absorption spectrum as given by the combination of Eqs. (2.11) and (3.16) for the same CdSe nanocrystal described above. In the strong confinement limit we find that the lowest transition observed is to the $1S_e - 1P_h$ exciton state at 2.7 eV, followed by the transition to the $1P_e - 1S_h$ exciton state at 3.1 eV and another transition to the $1P_e - 1D_h$ exciton state at 3.5 eV. The weak confinement limit shown in the lower panel of Figure 3 is also compared to the approximate result given by Eq. (3.18).

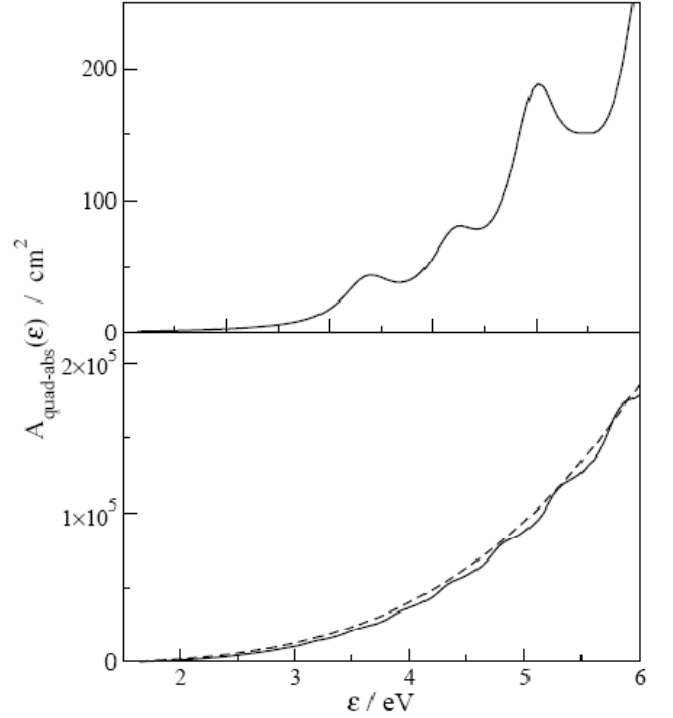

Figure 3: The quadrupole absorption cross section of a 2 nm (upper panel) and 5 nm (lower panel) CdSe nanocrystal. Effective masses (in atomic units): $m_e = 0.13$ and $m_h = 0.45$, $E_g = 1.7\ eV$, and $\varepsilon_{QD} = 10$. The broadening parameter is $\eta = 0.15$ eV. The dashed line in the lower panel is the weak confinement approximation given by Eq. (3.14) combined with (2.11).

In the weak confinement limit we find that the quadrupole oscillator strength is much larger than the corresponding dipole oscillator strength and is QD-size dependent. This can be attributed to the difference in the way the matrix elements depend on the QD radius, $R_{QD}$. The dipole transition moment is proportional to the overlap of the envelop functions, which is independent of $R_{QD}$, while the transition quadrupole moment is proportional to the envelop transition dipole $d_{env}$ (see Eq. (3.12)) which scales linearly with $R_{QD}$. The absorption spectra depend on the square of these transition moments and therefore, the quadrupole oscillator strength is larger by about a factor of $R_{QD}^2$.

## IV. FRET RATE BETWEEN QUANTUM DOTS

Explicit expressions for the FRET rate between QDs, involving the d-d and d-q contributions can be derived now. The donor-related quantities will be denoted by index *D* and those of the acceptor by *A*. The radii of the QDs is $R_{QD,i}$ ($i = D, A$), respectively and the average diameter is $D = R_{QD,A} + R_{QD,D}$. The donor emits from the lowest excitonic state in this model, which is the $1S_e - 1S_h$ exciton with energy

$$hc\bar{\nu}_{0,D} = E_{g,D} + \frac{\hbar^2 \pi^2}{2\mu_D R_{QD,D}^2} - \frac{1.8e^2}{\varepsilon_{QD,D} R_{QD,D}} \quad (4.1)$$

For FRET, this requires that the lowest excitonic states of the acceptor are of lower energy than those of the donor. We further assume that the emission spectrum of the donor is relatively narrow around this transition. From Eq. (2.15) we can obtain:

$$W_{D \to A}^{d-d} = \frac{2}{3}\left(\frac{\phi_D}{\tau_D \left(2\pi\bar{\nu}_{0,D}\right)^3}\right) \times \frac{(2\pi)^3 C_7 f_{1,D}^2 f_{1,A}^2}{\varepsilon^2 \bar{\nu}_{0,D}} \frac{A_{dip-abs}\left(2\pi c\bar{\nu}_{0,D}\right)}{D^6 \left(1 + r/D\right)^6}, \quad (4.2)$$

where *r* is the distance from the surface of the donor to the surface of the acceptor. Note that in our model the quantity $\phi_D / \tau_D \left(2\pi\bar{\nu}_{0,D}\right)^3$ is independent of the QD size (Eq. (2.14)).

In the strong confinement, we use Eq. (3.13) combined with Eq. (2.11) for the acceptor to obtain the FRET rate. In the weak confinement limit for the acceptor, ($R_{QD,A} \gg a_B$ where $a_B = \hbar^2 \varepsilon_{QD,A}/\mu_A e^2$ is the Bohr radius for the electron-hole pair of the acceptor), we use equation (3.14) and obtain:

$$W_{D \to A}^{d-d} = \frac{2}{3}\left(\frac{\phi_D}{\tau_D \left(2\pi\bar{\nu}_{0,D}\right)^3}\right) \frac{\mu_A d_{B,A}^2 f_{1,D}^2 f_{1,A}^2}{\hbar^2 \varepsilon^2} \frac{R_{QD,A}^3}{D^6 \left(1 + r/D\right)^6} \times \sqrt{\frac{2\mu_A}{\hbar^2}\left(hc\bar{\nu}_{0,D} - E_{g,A} + \frac{1.8e^2}{\varepsilon_{QD,A} R_{QD,A}}\right)}$$

(4.3)

In the case where the two QDs differ only with respect to the size, one can simplify the above equation. The square root becomes $\sqrt{\pi^2/R_{QD,D}^2 + 2\mu_A 1.8e^2/\hbar^2 \varepsilon_{QD}\left(R_{QD,D}^{-1} - R_{QD,A}^{-1}\right)}$. Assuming that one can neglect the second term relative to the confinement term, we find it is equal to $\pi/R_{QD,D}$, so that:

$$W_{D \to A}^{d-d} = \frac{2}{3}\left(\frac{\phi_D}{\tau_D \left(2\pi\bar{\nu}_{0,D}\right)^3}\right) \times \frac{\pi \mu_A d_{B,A}^2 f_{1,D}^2 f_{1,A}^2}{\hbar^2 \varepsilon^2} \frac{R_{QD,D}^{-1} R_{QD,A}^3}{D^6 \left(1 + r/D\right)^6}. \quad (4.4)$$

Finally we obtain the FRET rate in the weak confinement limit when the two QDs are made of the same material (but possibly different radii):

$$W_{D \to A}^{d-d} = \frac{2}{3}\left(\frac{\phi_D}{\tau_D \left(2\pi\bar{\nu}_{0,D}\right)^3}\right)^2 \frac{\pi \mu f_1^4}{4\hbar\varepsilon^2} \frac{R_{QD,D}^{-1} R_{QD,A}^3}{D^6 \left(1 + r/D\right)^6} \quad (4.5)$$

(same material)

We now consider the FRET rate due to dipole-quadrupole coupling, using the same assumptions as for the dipole-dipole coupling for the donor. From Eq. (2.15) we can obtain:

$$W_{D \to A}^{d-q} = 4\left(\frac{\phi_D}{\tau_D \bar{\nu}_{0,D}^3}\right) \frac{C_9 f_{1,D}^2 f_{2,A}^2}{\varepsilon^2 \bar{\nu}_{0,D}^3} \frac{A_{quad-abs}\left(2\pi c\bar{\nu}_{0,D}\right)}{D^8 \left(1 + r/D\right)^8} \quad (4.6)$$

In the strong confinement limit the quadrupole absorption cross-section is given in Eqs (2.11) and (3.16). Simplification can be obtained if the acceptor is weakly confined. Then Eq. (3.18) can be used to obtain:

$$W_{D \to A}^{d-q} = 4\left(\frac{\phi_D}{\tau_D \bar{\nu}_{0,D}^3}\right) \frac{\mu_A f_{1,D}^2 f_{2,A}^2 d_B^2}{4(2\pi)^3 \hbar^2 \varepsilon^2} \frac{R_{QD,A}^5}{D^8 \left(1 + r/D\right)^8} \times \sqrt{\frac{2\mu_A}{\hbar^2}\left(hc\bar{\nu}_{0,D} - E_{g,A} + \frac{1.8e^2}{\varepsilon_{QD,A} R_{QD,A}}\right)}$$

(4.7)

As analyzed for the dipole-dipole term, when the two QDs are made of the same material with different sizes Eq. (4.7) becomes:

$$W_{D \to A}^{d-q} = 4\left(\frac{\phi_D}{\tau_D \left(2\pi\bar{\nu}_{0,D}\right)^3}\right)^2 \frac{\pi \mu f_1^2 f_2^2}{16\hbar\varepsilon^2} \frac{R_{QD,D}^{-1} R_{QD,A}^5}{D^8 \left(1 + r/D\right)^8} \quad (4.8)$$

(same material)

## V. APPLICATIONS

We now discuss the specific applications of the theory developed above. We calculate the FRET rate in the usual dipole-dipole approximation and then consider the correction due to dipole-quadrupole coupling. Furthermore, within the simple effective mass approximation adopted here for the nanocrystal electronic structure, the fluorescence from ground excitonic state is quadrupole forbidden (dipole allowed) and thus, the quadrupole energy transition is only considered for the nanocrystal acceptor.

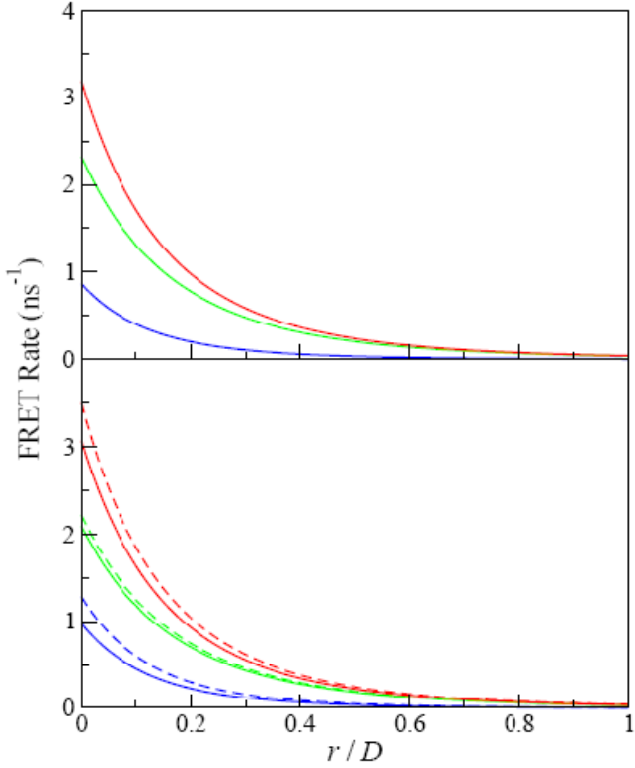

Figure 4: The FRET rate for coupling two CdSe QDs, $R_{QD,A} = 3nm$ (upper panel), $R_{QD,A} = 5nm$ (lower panel) and $R_{QD,D} = 2nm$. The dielectric constant of the medium is $\varepsilon = 1$. The total (red), dipole-dipole (green) and dipole-quadrupole (blue) FRET rates are shown. The remaining parameters are the same as in Figure 2. In the lower panel we show the exact result (solid) and the weak confinement approximation (dashed) for each case.

In Figure 4, we plot the FRET rate between two CdSe QDs in the strong and weak confinement limits for the acceptor QD. We observe that in both cases even at contact the contribution of the dipole-dipole (d-d) term is larger than that of the dipole-quadrupole (d-q). However, the latter is not negligible at contact and decays faster as the separation $r$ is increased. In the results shown here we assumed that the dielectric constant of the surrounding medium is $\varepsilon = 1$. $\varepsilon$ affects the FRET rate mostly as a scaling factor ($\varepsilon^{-2}$) as evident from Eqs. (4.4) and (4.6) although there is also a weak dependence of $f_\ell$ on $\varepsilon$. Thus, if experiments are done in organic solvents $(\varepsilon \approx 2)$, the FRET rates will be slower by a factor of 4, and in water $(\varepsilon \approx 80)$ the rates will be slower by nearly 4 orders of magnitude compared with the results shown in Figure 4.

The dependence of the FRET rates on the size of the QDs at the contact limit (the two QDs are nearly touching) is studied in Figure 5 (note that the acceptor radius must be larger than that of the donor, in order for the donor emission line to overlap the absorption spectrum of the acceptor). The most important features are: (a) the d-d contribution to the FRET rate (Eq. (4.2)) is larger than the d-q term (Eq. (4.8)) for all cases studied; (b) that the FRET rate decreases sharply as the donor size increases since the separation between the QD centers about which the multipolar expansion is carried out, increases.

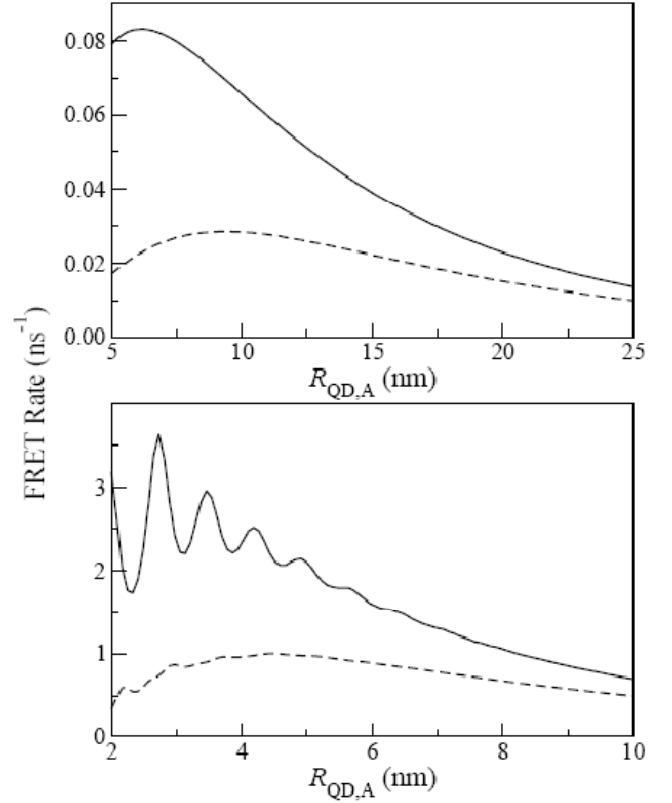

Figure 5: The d-d and d-q contributions to the FRET rate between a pair of CdSe QDs at contact limit as a function of the acceptor radius for two donor sizes: 5nm (top) and 2 nm (bottom). The remaining parameters are the same as in Figure 4.

As the size of the acceptor increases the contribution of the d-d term becomes comparable to the d-q term. This can be analyzed within the weak confinement limit where one can derive a simple relation between the d–d and the d-q contributions to the rate (see Eqs. (4.5) and (4.8)):

$$\frac{W_{D\to A}^{d-d}}{W_{D\to A}^{d-q}} = \frac{6}{25}\left(\frac{2\varepsilon_{QD} + 3\varepsilon}{\varepsilon_{QD} + 2\varepsilon}\right)^2 \left(\frac{D+r}{R_{QD,A}}\right)^2 \quad (5.1)$$

This ratio depends weakly on $\varepsilon_{QD}$ and more pronouncedly on the radii of the two QDs. It does not depend on the band gap nor on the effectives masses of the QDs, thus the ratio is expected to be a universal quantity. Analyzing this relation, we observe that at contact the ratio varies between 0.5 (when $\varepsilon \gg \varepsilon_{QD}$) and close to 1 (when $\varepsilon \ll \varepsilon_{QD}$), for a large acceptor, as indeed is observed in Figure 5. In addition, as the separation $r$ is increased, we find from Eq. (5.1) that the d-d term becomes more dominant.

An interesting feature of small QDs is the existence of a structure in the d-d and d-q contributions to the FRET rate as a function of the acceptor size. The peaks correspond to resonances between the emission lines of the donor and absorption lines of the acceptor, whose positions vary with the QD size. The structure is washed out as the acceptor approaches its weak confinement limit due to the larger density of states at energies corresponding to the emitting donor. As can be seen in the figure, the structure is considerably more pronounced in the d-d than in the d-q contributions. This can be traced to the more structured dipole absorption spectra (Figure 2) compared to that of the quadrupole (Figure 3).

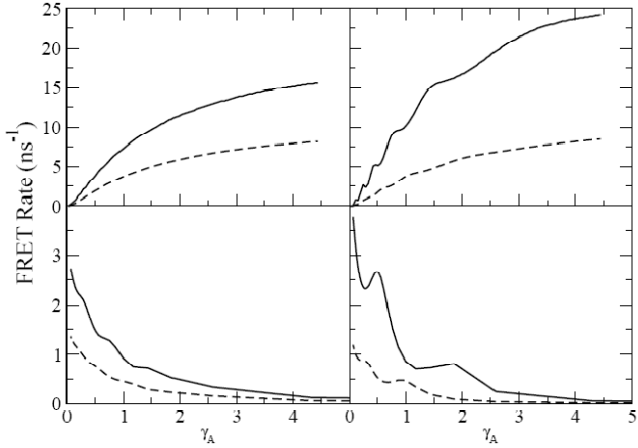

Figure 6: The d-d (solid) and d-q (dashed) FRET rates as a function of the acceptor mass ratio $\gamma_A = m_e/m_h$. Right panels are for $R_{QD,A} = 3\,\text{nm}$ and left panels are for $R_{QD,A} = 5\,\text{nm}$. In the lower panels we keep $m_{e,A} = 0.13$ and vary $m_{h,A}$. In the upper panels we keep $m_{h,A} = 0.45$ and vary $m_{e,A}$. The remaining parameters are identical to those shown in Figure 4 for QD separation $r = 0\,\text{nm}$.

In Figure 6, we study the effect of the electron and hole mass ratios of the acceptor on the FRET rate. Physically this parameter is not adjustable (although one can affect it by changing the QD material). However, it is instructive to determine the way it can potentially affect the rate. The results shown in Figure 6 are for CdSe nanocrystals with fictitious electron and hole masses. We modify the mass ratio by either changing $m_{h,A}$ holding $m_{e,A}$ fixed, or vice versa. The effective mass changes in each of the two cases, according to the formula:

$$\mu_A = m_{h,A} \frac{1}{1+\gamma_A^{-1}} = m_{e,A} \frac{1}{1+\gamma_A} \qquad (5.2)$$

When the electron mass is kept constant, the increase of $\gamma_A$ causes a *decrease* of the effective mass and the FRET rate decreases (because the density of states decreases due to increased confinement). When the mass of the hole is held constant the effective mass *increases* with the growing $\gamma_A$, and the FRET rate grows. Comparing the results, we find that the overall FRET rate is larger when the electron mass is varied. This is due to the fact that the corresponding mass of the hole is relatively large, giving rise to a smaller confinement effects. While, for the case that the hole mass is varied, the corresponding electron mass is small and thus, due to the quantum confinement, very few transitions overlap the donor emission line. Comparing the results for different acceptor sizes, we find that when the acceptor QD is small, the FRET rate is characterized by a resonant structure, signifying once again the resonances between the emission line and the absorption. As the effective mass is varied different absorption lines of the acceptor enter the emission window of the donor.

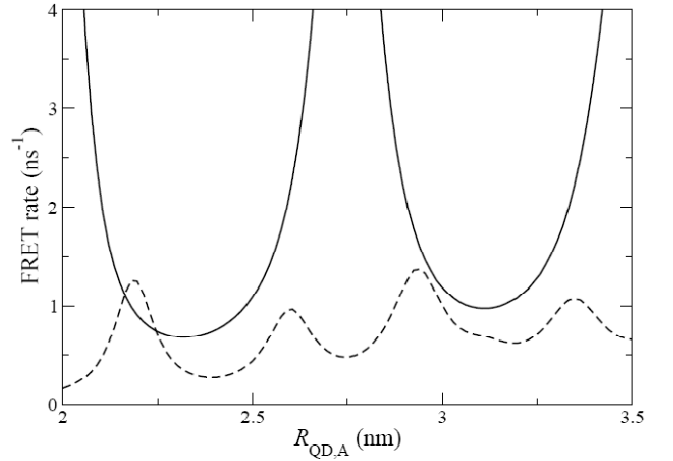

Figure 7: The d-d (solid) and d-q (dashed) FRET rates for the coupling between two CdSe QDs, with $R_{QD,D} = 2$ nm and the acceptor size is variable. The line width parameter for computing the absorption spectrum is $\eta = 0.05$ eV. The remaining parameters are the same as in Figure 2.

An interesting question is whether the d-d contribution to the FRET rate is always larger than the d-q contribution. Since the selection rules for absorption within the dipole and quadrupole approximations are different, one can envision a situation where the donor emission line aligns with a d-q allowed transition and much less so with a d-d allowed one for a certain acceptor size. For realistic broadenings this does not usually happen. However, this becomes possible if one considers much narrower absorption line widths, a situation that might be achievable when appropriate size selection and surface control are achieved. In Figure 7 we show an example of such occurrence. We considered two CdSe QDs with a donor of radius 2 nm. When the width parameter is small enough, namely $\eta = 0.05$ eV in Eq. (3.15) is used, we find that acceptor sizes around a relatively narrow region slightly above 2 nm the d-q contribution exceeds that of the d-d. We can explain the behavior seen in the figure as follows. When the two dots are of the same size (2 nm) the d-d contribution is large because both have the $1S_e - 1S_h$ exciton transition in full resonance. However, as the acceptor size grows, its $1S_e - 1S_h$ exciton line sharply falls in energy and out of resonance with that of the donor. As a result, the d-d coupling diminishes significantly. As the size of the acceptor grows further, its $1S_e - 1P_h$ exciton line also lowers and eventually enters into resonance with that of the donor $1S_e - 1S_h$ line. This causes a strong d-q coupling which can slightly exceeds that of the d-d coupling as seen in the figure.

## VI. SUMMARY AND DISCUSSION

In this article we have studied the FRET rate between a pair of semiconducting nanocrystal QDs. We considered two types of donor-acceptor couplings for the FRET rate: the dipole-dipole (d-d) and the dipole-quadrupole (d-q) coupling. Using an extended Förster theory given by Eq. (2.15), we derived equations for the d-d (Eq. (4.2)) and d-q (Eq. (4.6)) FRET rate between QDs as a function of their separation. This was derived assuming that the donor emission line is narrow. All one needs to know in order to use these equations is the size of the QDs, the dipole and quadrupole absorption spectra of the acceptor, the radiative lifetime and the lowest exciton energy of the donor, the dielectric constants of the QDs and the dielectric constant of the surrounding medium.

In order to use these equations in a theoretical setting, we have adopted the effective mass model to derive expressions for the measureable quantities as a function of the QD distance and in terms of their physical properties (size, fundamental gap, effective mass, exciton radius and dielectric constant). This allowed us to delineate the various factors that govern the FRET rate. We have also considered the weak confinement limit, where the expressions can be further simplified and the FRET rates are given by Eqs. (4.5) (d-d contribution) and (4.8) (d-q contribution). Our major conclusions are summarized as follows:

1) The d-d contribution to the FRET rate decays as $R^{-6}$ (where $R$ is the QD center-center separation) and is almost always larger than that of the d-q contribution which decays faster (decays as $R^{-8}$). Nevertheless, the latter is not negligible at short QD separations and must be taken into account for a quantitative description.
2) The difference in the scaling of the d-d and d-q contributions to the FRET rate is quite difficult to detect experimentally. Thus, in many cases, an "effective" d-d term can be used to describe the overall FRET rate.
3) In certain cases, when the spectral lines are narrow, the d-q contribution to the FRET rate may become even larger than the d-d contribution for certain QD sizes. This effect is due to sharp resonances.
4) In small QDs (size small compared to their exciton Bohr radius) we find strong dependence of the FRET rate on the size. This is caused by the sensitivity of the spectral overlaps to the confinement.
5) The effect of the dielectric constant of the dots is relatively weak compared to the effect of the dielectric medium of the surrounding on the FRET rate.
6) The effective masses can affect the FRET rates considerably, mainly by changing the density of states (i.e. the spectral overlap).

We believe that the results presented in this work provide a qualitative picture of the FRET behavior of nanocrystal QDs. Future work will attempt to extend the theory in several directions so that more quantitative features can be addressed. More accurate treatments of the electronic structure can be used, such as the finite confining potential, including more than two bands and coupling between the different bands etc. In addition, we will employ an atomistic description of the QDs electronic structure based on a semi-empirical method.

## ACKNOWLEDGMENTS

We thank Professors Uri Banin and Itamar Willner for useful discussions. This research was supported by the Converging Technologies Program of The Israel Science Foundation (grant number 1704/07) and The Israel Science Foundation (grant number 962/06).

## APPENDIX A: LOCAL FIELD CORRECTION FACTOR

Let us consider a point charge $q$ located at position **s** inside a sphere (QD) of dielectric constant $\varepsilon_{QD}$ of radius $R_{QD}$ surrounded by a medium of dielectric constant $\varepsilon$. The electric potential at position **r** *outside* of the QD is:[36]

$$V(\mathbf{r};\mathbf{s}) = \frac{q}{\varepsilon_{QD}} \left( \frac{1}{|\mathbf{r}-\mathbf{s}|} + \frac{1}{r}\sum_{\ell=0}^{\infty} g_l \left(\frac{s}{r}\right)^{\ell} P_{\ell}(\cos\theta) \right), \quad (A.1)$$

where:

$$g_{\ell} = \frac{(\ell+1)\left(\frac{\varepsilon_{QD}}{\varepsilon}-1\right)}{\ell\left(\frac{\varepsilon_{QD}}{\varepsilon}+1\right)+1} \quad (A.2)$$

This potential includes a Coulomb term and a correction term, accounting for $\varepsilon \neq \varepsilon_{QD}$. Expanding the Coulomb term in a multipole series, $\frac{1}{|\mathbf{r}-\mathbf{s}|} = \frac{q}{r}\sum_{\ell=0}^{\infty}\left(\frac{s}{r}\right)^{\ell} P_{\ell}(\cos\theta)$, combining it with the correction series:

$$V(\mathbf{r};\mathbf{s}) = \frac{q}{\varepsilon r}\sum_{\ell=0}^{\infty} f_l\left(\frac{\varepsilon}{\varepsilon_{QD}}\right)\left(\frac{s}{r}\right)^{\ell} P_{\ell}(\cos\theta), \quad r > R_{QD}. \quad (A.3)$$

The effect of the QD is to augment the $\ell^{th}$ pole by a local-field factor

$$f_\ell(x) = \frac{2\ell + 1}{(x+1)\ell + 1}. \tag{A.4}$$

Interestingly, aside from the requirement that $r > R_{QD}$, the result, Eq. (A.3) is not dependent explicitly on the size of the QD. Now consider a dipole ($\ell = 1$), a quadrupole ($\ell = 2$), or any multipole of order $\ell$ located in the center of the QD. It is easily seen that Eq. (A.3) implies that outside the dot, the electric potential field is that of a multipole $\ell$ in a homogeneous medium $\varepsilon$. The effect of the QD is solely in screening (or descreening) the multipole moment by the field factor $f_\ell(\varepsilon_{QD}/\varepsilon)$. This result simplifies considerably the treatment of the screening effects.

Now consider the case where we have two dipoles inside two different QDs (with dielectric constants $\varepsilon_{QD,1}$ and $\varepsilon_{QD,2}$) embedded in a medium of dielectric constant $\varepsilon$. In principle, one must solve the corresponding Poisson equation with the proper boundary conditions. The discussion above however suggests the following reasonable and *simple* approximation: that the only effect of the dielectric constants is in its screening the corresponding multipoles. Thus, under this approximation the dipole-dipole coupling (Eq. (2.3)) needs only be is multiplied by the two local field factors:

$$V_{if} \to V_{if} f_1(\varepsilon_{QD,1}/\varepsilon) f_1(\varepsilon_{QD,2}/\varepsilon), \tag{A.5}$$

As for the dipole-quadrupole coupling (Eq. (2.4)), the same reasoning leads to:

$$V_{if} \to V_{if} f_2(\varepsilon_{QD,1}/\varepsilon) f_2(\varepsilon_{QD,2}/\varepsilon), \tag{A.6}$$

## APPENDIX B: ORIENTATION AVERAGING

Averaging of the FRET rate over random orientations requires computing the average of the square of the coupling matrix $|V_{if}|^2$. For the dipole-dipole coupling we have, using Eq (2.3):

$$\begin{aligned}\left\langle |V_{if}^{d-d}|^2 \right\rangle &= C \left\langle (\cos\theta_{\alpha\delta} - 3\cos\theta_\alpha \cos\theta_\delta)^2 \right\rangle \\ &= C \{ \langle \cos^2\theta_{\alpha\delta} \rangle - 6\langle \cos\theta_{\alpha\delta} \cos\theta_\alpha \cos\theta_\delta \rangle + \\ & \quad 9 \langle \cos^2\theta_\alpha \rangle \langle \cos^2\theta_\delta \rangle \} \end{aligned} \tag{B.1}$$

where $C = \left(\dfrac{d^\delta d^\alpha}{\varepsilon R^3}\right)^2$. Using the following averages:

$$\begin{aligned} \langle \sin\theta \rangle &= \frac{\pi}{4} & \langle \cos\theta \rangle &= 0 \\ \langle \cos^2\phi \rangle &= \frac{1}{2} & \langle \sin^2\phi \rangle &= \frac{1}{2} \\ \langle \sin^2\theta \rangle &= \frac{2}{3} & \langle \cos^2\theta \rangle &= \frac{1}{3} \\ \langle \cos^4\theta \rangle &= \frac{1}{5} & \langle \sin^2 2\theta \rangle &= \frac{8}{15} \end{aligned} \tag{B.2}$$

We obtain:

$$\begin{aligned} \langle \cos^2\theta_{\alpha\delta} \rangle &= \frac{1}{3} \\ \langle \cos\theta_{\alpha\delta} \cos\theta_\alpha \cos\theta_\delta \rangle &= \frac{1}{9} \end{aligned} \tag{B.3}$$

And thus evaluate:

$$\left\langle |V_{if}^{d-d}|^2 \right\rangle = \frac{2}{3}\left(\frac{d^\delta d^\alpha}{\varepsilon R^3}\right)^2 \tag{B.4}$$

As for averaging the dipole-quadrupole interaction in Eq. (2.4), using the same reasoning and Eqs. (B.2) it is straightforward to obtain:

$$\left\langle |V_{if}^{d-q}|^2 \right\rangle = \left(\frac{2d^\delta \Theta^\alpha}{\varepsilon R^4}\right)^2. \tag{B.5}$$

## APPENDIX C: THE ENVELOPE TRANSITION DIPOLE MOMENT

We consider the envelope transition dipole moment:

$$\begin{aligned}\left\langle \phi_{h,nlm} | z | \phi_{e,n'l'm'} \right\rangle &= N_{n,l} N_{n,l'} \\ &\int_0^{R_{QD}} r^3 j_l\left(\kappa_{n,l} \frac{r}{R_{QD}}\right) j_{l'}\left(\kappa_{n',l'} \frac{r}{R_{QD}}\right) dr \\ &\int \sin\theta d\theta d\phi \cos\theta Y_{lm}(\theta,\phi) Y_{l'm'}(\theta,\phi) \end{aligned} \tag{C.1}$$

Where:

$$N_{n,l} = \sqrt{\frac{2}{R^3}}(j_{l+1}(\kappa))^{-1} \tag{C.2}$$

is the normalization constant. And the spherical harmonics are:

$$\begin{aligned} Y_{lm}(\theta,\phi) &= \sqrt{\frac{1}{2\pi \tilde{N}_{lm}}} P_l^m(\cos\theta) e^{im\phi} \\ \tilde{N}_{lm} &= \frac{2}{2l+1} \frac{(l+m)!}{(l-m)!} \end{aligned} \tag{C.3}$$

where $P_l^m(x)$ are the associated Legendre functions. The angular part gives the selection rules: $l \to l' \pm 1$ and $m \to m' \pm 1$. For the z component the $m = m'$ condition prevails and using the following relation between Legendre functions[43]

$$(l+1-m)P_{l+1}^m(x) + (l+m)P_{l-1}^m(x) = (2l+1)xP_l^m$$

$$\frac{1}{\sqrt{\tilde{N}_{lm}\tilde{N}_{l'm}}} \int_{-1}^{1} P_l^m(x) P_{l'}^m(x) dx = \delta_{ll'} \quad \text{(C.4)}$$

we obtain after some algebra:

$$\int_{-1}^{1} dx\, x P_l^m(x) P_{l'}^m(x) = F_{l'}^m \delta_{l+1,l'} + F_l^m \delta_{l,l'+1} \quad \text{(C.5)}$$

Where:

$$F_l^m = \sqrt{\frac{(l+m)(l-m)}{(2l+1)(2l-1)}} \quad \text{(C.6)}$$

And for the radial part:

$$N_{n,l} N_{n',l'} \int_0^{R_{QD}} r^3 j_l\left(\kappa_{n,l}\frac{r}{R_{QD}}\right) j_{l'}\left(\kappa_{n',l'}\frac{r}{R_{QD}}\right) dr$$

$$= \frac{2 R_{QD}}{j_{l+1}(\kappa_{nl}) j_{l'+1}(\kappa_{n'l'})} \int_0^1 x^3 j_l(\kappa_{n,l} x) j_{l'}(\kappa_{n',l'} x) dx \quad \text{(C.7)}$$

$$\approx \frac{R_{QD}}{6} \begin{cases} 3 & n'=n, l'=l\pm 1 \\ 2 & n'=n\mp 1, l'=l\pm 1 \\ 0 & \text{otherwise} \end{cases}$$

The last part of the expression is an approximation to the exact numerical result. We find that elements with $|n-n'|>1$ are smaller by an order of magnitude or more, and can be neglected.

## REFERENCES

[1] X. S. Xie and R. C. Dunn, Science **265** (5170), 361 (1994).
[2] W. E. Moerner and M. Orrit, Science **283** (5408), 1670 (1999).
[3] S. Weiss, Science **283** (5408), 1676 (1999).
[4] P. Alivisatos, Nat Biotechnol **22** (1), 47 (2004).
[5] A. P. Alivisatos, W. W. Gu, and C. Larabell, Annu. Rev. Biomed. Eng. **7**, 55 (2005).
[6] I. Willner, R. Baron, and B. Willner, Biosensors & Bioelectronics **22** (9-10), 1841 (2007).
[7] H. Mattoussi, J. M. Mauro, E. R. Goldman, G. P. Anderson, V. C. Sundar, F. V. Mikulec, and M. G. Bawendi, J. Am. Chem. Soc. **122** (49), 12142 (2000).
[8] A. R. Clapp, I. L. Medintz, J. M. Mauro, B. R. Fisher, M. G. Bawendi, and H. Mattoussi, J. Am. Chem. Soc. **126** (1), 301 (2004).
[9] E. Katz and I. Willner, Angew. Chem. Int. Edit. **43** (45), 6042 (2004).
[10] R. Gill, I. Willner, I. Shweky, and U. Banin, J. Phys. Chem. B **109** (49), 23715 (2005).
[11] X. H. Gao and S. M. Nie, Trends in Biotechnology **21** (9), 371 (2003).
[12] W. C. W. Chan, D. J. Maxwell, X. H. Gao, R. E. Bailey, M. Y. Han, and S. M. Nie, Current Opinion in Biotechnology **13** (1), 40 (2002).
[13] X. Michalet, F. Pinaud, T. D. Lacoste, M. Dahan, M. P. Bruchez, A. P. Alivisatos, and S. Weiss, Single Molecules **2** (4), 261 (2001).
[14] A. J. Sutherland, Current Opinion in Solid State & Materials Science **6** (4), 365 (2002).
[15] T. Förster, Discuss. Faraday Soc. **27**, 7 (1959).
[16] G. D. Scholes, Ann. Rev. Phys. Chem. **54**, 57 (2003).
[17] G. D. Scholes and D. L. Andrews, Phys. Rev. B **72** (12) (2005).
[18] G. Parascandolo and V. Savona, Phys. Rev. B **71** (4), 045335 (2005).
[19] A. L. Efros, M. Rosen, M. Kuno, M. Nirmal, D. J. Norris, and M. Bawendi, Phys. Rev. B **54** (7), 4843 (1996).
[20] A. L. Efros and M. Rosen, Annu Rev Mater Sci **30**, 475 (2000).
[21] A. Zunger, Phys Status Solidi B **224** (3), 727 (2001).
[22] A. Franceschetti and A. Zunger, Phys. Rev. B **62** (4), 2614 (2000).
[23] H. X. Fu, L. W. Wang, and A. Zunger, Appl. Phys. Lett. **71** (23), 3433 (1997).
[24] H. X. Fu and A. Zunger, Phys. Rev. B **56** (3), 1496 (1997).
[25] L. W. Wang and A. Zunger, Phys. Rev. B **53** (15), 9579 (1996).
[26] E. Rabani, B. Hetenyi, B. J. Berne, and L. E. Brus, J. Chem. Phys. **110** (11), 5355 (1999).
[27] D. L. Dexter, J. Chem. Phys. **21** (5), 836 (1953).
[28] G. D. Scholes and D. L. Andrews, J. Chem. Phys. **107** (14), 5374 (1997).
[29] A. Salam, Int. J. Quant. Chem. **105** (6), 762 (2005).
[30] A. L. Efros and A. L. Efros, Sov. Phys. - Semicond. **16** (7), 772 (1982).
[31] L. E. Brus, J. Chem. Phys. **79** (11), 5566 (1983).
[32] L. E. Brus, J. Chem. Phys. **80** (9), 4403 (1984).
[33] Y. Z. Hu, M. Lindberg, and S. W. Koch, Phys. Rev. B **42** (3), 1713 (1990).
[34] A. I. Ekimov, A. L. Efros, and A. A. Onushchenko, Sol. Stat. Comm. **56** (11), 921 (1985).
[35] H. Weller, H. M. Schmidt, U. Koch, A. Fojtik, S. Baral, A. Henglein, W. Kunath, K. Weiss, and E. Dieman, Chem. Phys. Lett. **124** (6), 557 (1986).
[36] L. Bányai and S. W. Koch, *Semiconductor quantum dots*. (World Scientific, Singapore ; River Edge, NJ, 1993).
[37] D. B. T. Thoai, Y. Z. Hu, and S. W. Koch, Phys. Rev. B **42** (17), 11261 (1990).
[38] D. J. Norris, A. L. Efros, M. Rosen, and M. G. Bawendi, Phys. Rev. B **53** (24), 16347 (1996).
[39] A. L. Efros, Phys. Rev. B **46** (12), 7448 (1992).
[40] C. A. Leatherdale, W. K. Woo, F. V. Mikulec, and M. G. Bawendi, J. Phys. Chem. B **106** (31), 7619 (2002).
[41] W. W. Yu, L. H. Qu, W. Z. Guo, and X. G. Peng, Chem. Mater. **15** (14), 2854 (2003).
[42] U. Banin and O. Millo, Ann. Rev. Phys. Chem. **54**, 465 (2003).
[43] M. Abramowitz and I. A. Stegun, *Handbook of Mathematical Functions with Formulas, Graphs, and Mathematical Tables*, 10 ed. ( U.S. Department of Commerce, Washington D.C., 1972).